\documentclass[pre,twocolumn,groupedaddress,showpacs]{revtex4}
\usepackage{graphicx}
\usepackage{dcolumn}
\usepackage{amsmath}
\usepackage{bm}

\newcommand{\bfn}{{\bf n}}
\newcommand{\bfr}{{\bf r}}
\newcommand{\bfQ}{{\bf Q}}
\newcommand{\bfS}{{\bf S}}
\newcommand{\bfM}{{\bf M}}
\newcommand{\bfu}{{\bf u}}
\newcommand{\bfe}{{\bf e}}
\newcommand{\bfI}{{\bf I}}
\newcommand{\rmd}{{\rm d}}
\newcommand{\rmTr}{{\rm Tr}}
\newcommand{\calR}{{\cal R}}
\begin{document}

\title{Molecular simulation and theory of a liquid crystalline disclination core} 
\author{Denis Andrienko} 
\author{Michael P. Allen}
\affiliation{H. H. Wills Physics Laboratory, University of Bristol, Bristol BS8 1TL, United Kingdom
}

\begin{abstract}
Molecular simulations of a nematic liquid crystal confined in cylinder geometry
with homeotropic anchoring have been carried out.
The core structure of a disclination line defect of strength $+1$
has been examined, and comparison made with various theoretical treatments,
which are presented in a unified way.
It is found that excellent fits to the cylindrically-symmetrized
order tensor profiles may be obtained with appropriate parameter choices;
notwithstanding this, on the timescales of the simulation,
the cylindrical symmetry of the core is broken and two
defects of strength $+1/2$ may be resolved.
\end{abstract}
\pacs{PACS: 61.30.Cz,61.30.Jf,61.20.Ja,07.05.Tp}
\maketitle

\section{Introduction}
\label{sec:intro}
Liquid crystals are at the heart of a range of technological devices,
which rely on the ability to manipulate 
the direction of preferred molecular alignment,
the \emph{director} $\bfn$,
with electric and magnetic fields,
and through coupling to surfaces.
The practical performance of such devices,
and the theoretical description of director distortions,
both rely on smooth variation of the director
with position in space;
this is well accounted for by the Frank free energy
\cite{frank.fc:1958.a,degennes.pg:1995.a}
which is an expansion in squared gradients
of the director field $\bfn(\bfr)$,
parametrized by
the splay ($K_{11}$), twist ($K_{22}$) and bend ($K_{33}$)
elastic constants.
The director $\bfn$ is the principal axis
of the local second-rank ordering tensor $\bfQ$,
which characterizes the nematic state;
such a description is intrinsically uniaxial,
and neglects variation of the \emph{degree} of ordering
with position,
of the kind which occurs near bounding surfaces
and around topological defects.
Such defects are treated as singularities in the director field;
for a better description, it is necessary to replace
the director field by one which
allows some variation of the relevant order tensor components
over relatively short length scales.
The relevant region near the defect is called the \emph{core}.

The most common type of defect in the nematic phase
is the disclination line defect,
characterized by an integer or half-integer index
defined by the number of turns of the director field
associated with taking a circuit about the line.
The phenomenological approach to the investigation of line defects was 
described by Schopohl and Sluckin 
\cite{schopohl.n:1987.a,schopohl.n:1988.a}.
They worked with 
the full order-parameter tensor $Q_{\alpha\beta}$ 
in the framework of Landau-de Gennes theory,
and applied this to the
structure of the $\pm\frac{1}{2}$  disclination core.
The \emph{full} order parameter $Q_{\alpha \beta } $ was used 
a) to avoid \emph{divergent} terms in the elastic energy; 
b) to take into account possible \emph{biaxiality} of the defect core.
Then, the general idea was implemented for the particular problem of the
core structure of the $\pm 1$ strength defect: Biscari and Virga 
\cite{biscari.p:1997.a} 
and Mottram and Hogan 
\cite{mottram.nj:1997.a} solved the equations
for the order tensor and obtained order parameter and biaxiality profiles.
They used truncated expansions of the free energy density, which helped them
to obtain analytical solutions to the problem.
Finally, Sigillo et al 
\cite{sigillo.i:1998.a} 
considered the $+1$ disclination problem in the spirit
of Maier-Saupe mean-field theory.

In this paper we use the approach proposed by Schopohl and Sluckin to obtain
the order parameter profiles near the disclination line. We also
recapitulate the results of the other theories in order to discuss the
advantages and disadvantages of the proposed models.
We compare these predictions with the results of
computer simulation using a simple molecular model.
Computer simulation is a well-established method of relating
bulk elastic coefficients
\cite{1988.a,1990.a,1992.a,1996.g,stelzer.j:1995.a,stelzer.j:1995.b,stelzer.j:1997.a}
and more recently surface anchoring strengths
\cite{stelzer.j:1997.b,1999.b},
to molecular structure and interaction parameters.
An early study of disclination line defects
\cite{hudson.sd:1993.a}
involved Monte Carlo simulations of rod-like molecules
(hard spherocylinders)
confined in cylindrical geometry
with boundary conditions chosen to stabilize the chosen director field
far from the defect.
This work concentrated on the disclination line defect of strength 
$-\frac{1}{2}$, in cylinders of radius 2--3 times the molecular length.
For short rods, smooth variation of the order parameters with position
was observed,
and the defect core retained axial symmetry.
For longer rods, having much larger bend elastic constants,
this symmetry was broken, and various microscopically `escaped' structures
were seen.
There was also evidence of metastability,
and non-convergence of structures from different starting configurations,
which the authors attempted to resolve by free energy calculations.
More recently, a study of two-dimensional models has been carried out
\cite{Dzubiella:1999.a}.
These studies support the view that the disclination core
is one or two molecular lengths across.

The purpose of this paper is to present
simulation results for a model of hard particles in a cylindrical pore,
following closely the approach of Ref.~\cite{hudson.sd:1993.a},
and compare with the theories just mentioned.
We study the simplest example of biaxial molecular arrangement, 
namely the disclination defect of strength $+1$
corresponding to a uniform, cylindrically symmetric,
splay deformation of the director far from the core. 
This is imposed by choosing homeotropic anchoring conditions
at the cylinder pore walls.

We take care to check the equilibration of our simulations,
and note that quite extensive runs are required to ensure this.
We also pay particular attention to studying the effects of
pore size, with radius varying from 2 to 5 times the molecular length,
so as to eliminate the effects of the walls on the defect core.
More precisely, our aim is to restrict these effects to those which arise
from the planar radial far-field structure,
rather than from density and order parameter variations 
in the immediate vicinity of the walls.
It is well-known that the pore radius plays a critical role
in determining the stability of different nematic structures
in this geometry.
An analysis of the elastic free energy
shows that the `escaped radial' structure, in which the director
bends over to become parallel with the symmetry axis
in the core region, 
is the stable structure for large pore radius
\cite{cladis.pe:1972.a,kleman.m:1983.a}.
In analyzing experiments on nematic liquid crystals
in cylindrical pores,
Crawford et al.
\cite{crawford.gp:1991.a,crawford.gp:1992.a}
show that at small pore radii, 
the escaped structure (which will also, in general,
have point defects along the disclination line)
is not the most stable;
they discuss the planar radial and planar polar configurations,
finding the latter to be stable for the parameter values
that they survey.

A unified presentation of the different theories is given in
Sec.~\ref{sec:theory};
the model, and simulation techniques,
are set out in Sec.~\ref{sec:model}.
The observed structures
are described in Sec.~\ref{sec:results},
along with theoretical fits to
the order parameter profiles.
A discussion of the results,
and the validity of the theories, together with some concluding remarks,
appear in Sec.~\ref{sec:conclusions}.
\section{Phenomenological models}
\label{sec:theory}
In the framework of the continuum theory, 
the system can
be described by the Landau-de Gennes free energy density 
\cite{degennes.pg:1995.a}: 
\begin{equation}
F= \kappa \left| \nabla \bfQ\right| ^{2}
+\sigma \left(\bfQ\right) \:,  
\label{free_energy_full}
\end{equation}
where $\sigma\left(\bfQ\right)$ is a function of 
the invariants of $\bfQ$, 
the symmetric order tensor.
This is the suitably-normalized traceless part
of the tensor $\bfM$ of the second moments of the
molecular orientational distribution function $f$
\begin{eqnarray}
\bfM &=& \int ~ f\left(\bfu\right)~
  \bfu\otimes\bfu~\rmd\Omega , 
\label{eq_order_definition} 
\\
\bfQ &=& \frac{3}{2}\bfM-\frac{1}{2}\bfI \:,
\nonumber
\end{eqnarray}
where $\bfI$ is the unit tensor
and the integration is over the unit sphere.

We exclude the escape of the director in $z$ direction,
and our boundary conditions also prevent spiral configurations,
so the eigenvectors of $\bf Q$\ coincide with the 
unit vectors of the cylindrical geometry,
namely $\bfe_{\rho}$, $\bfe_{\theta}$ and $\bfe_{z}$,
respectively,
in the radial, tangential, and axial directions.
Since the eigenvalues of $\bfQ$\ are not independent, 
the tensor may be
expressed in terms of two independent parameters: the order parameter $S$
and biaxiality $\alpha$ 
\cite{biscari.p:1997.a}: 
\begin{eqnarray}
\nonumber
\bfQ &\equiv&
Q_{\rho\rho}~\bfe_{\rho}\otimes\bfe_{\rho } + 
Q_{\theta\theta}~\bfe_{\theta }\otimes\bfe_{\theta} +
Q_{zz}~\bfe_{z}\otimes\bfe_{z}
\\
\nonumber
&=& S~\bfe_{\rho}\otimes\bfe_{\rho } + 
\left(-\frac{1}{2}S+\frac{3}{2}\alpha \right)~
\bfe_{\theta }\otimes\bfe_{\theta} 
\\
&+& \left(-\frac{1}{2}S-\frac{3}{2}\alpha\right)~\bfe_{z}\otimes\bfe_{z} \:.
\label{eq_parametrization}
\end{eqnarray}

Minimizing the full free energy functional (\ref{free_energy_full}) with
respect to the $S, \alpha$ and taking into account that 
$S=S\left(\rho\right)$, $\alpha=\alpha\left(\rho\right)$,
where $\rho$ is the distance from the cylinder axis, 
we obtain the following Euler-Lagrange equations: 
\begin{eqnarray}
\label{general_equations}
\left( \rho S^{\prime }\right) ^{\prime }-3\rho ^{-1}\left( S-\alpha \right)
-3g_{1}\left( S,\alpha \right)  &=&0, \\
\left( \rho \alpha ^{\prime }\right) ^{\prime }+\rho ^{-1}\left( S-\alpha
\right) -g_{2}\left( S,\alpha \right)  &=&0.  \nonumber
\end{eqnarray}
Here $g_{1}\left( S,\alpha \right) =\frac{2}{9}\kappa^{-1}\partial \sigma /\partial S$, 
$g_{2}\left( S,\alpha \right) =\frac{2}{9}\kappa^{-1}\partial \sigma /\partial \alpha $,
and the prime denotes differentiation with respect to $\rho$.

The system of equations (\ref{general_equations}) should be solved with
corresponding boundary conditions. 
The boundary conditions at $\rho=0$ can be obtained if we look for the
solutions close to the centre of the disclination line. Indeed, in the limit 
$\rho \rightarrow 0$, we can neglect the functions $g_{1},g_{2}$ in
equations~(\ref{general_equations}). Then, seeking the solution as an
expansion in powers of $\rho$, one can obtain that in the region close to
the centre of the disclination line the solutions are 
\begin{equation}
S = S_{0}+3\gamma\rho ^{2} \:,   \qquad
\alpha  = S_{0}-\gamma \rho ^{2} \:,  
\label{solutions}
\end{equation}
where $S_{0}$ and $\gamma$ are constants.
Therefore, at $\rho =0$, we have 
\begin{equation}
\left. S^{^{\prime }}\right| _{\rho =0}=0 \:,
\qquad
\left. \alpha ^{^{\prime }}\right|_{\rho =0}=0   \:.
\label{boundary_0}
\end{equation}
We also assume that, far away from the disclination core, 
we have uniaxial nematic, 
and the surface at $\rho=R$ provides the order parameter $S_{s}$: 
\begin{equation}
\left. S\right| _{\rho =R}=S_{s} \:,
\qquad
\left. \alpha \right| _{\rho =R}=0 \:.
\label{boundary_R}
\end{equation}

Some general properties of the equations (\ref{general_equations}),
regardless of the explicit form of the functions $g_{1}$, $g_{2}$, 
can help us to fit the simulation results. 
The simulations provide the value of the
order parameter on the disclination line $S_{0}$ which we can also derive
analytically.
Indeed, the solutions (\ref{solutions}) imply that 
$\left. S+3\alpha \right|_{\rho \rightarrow 0}=\mbox{const}$. 
The condition that the energy is bounded
requires also $S_{0}=\alpha _{0}$. 
Therefore, from (\ref{general_equations})
we obtain the implicit equation for the order parameter value on the
disclination line: 
\begin{equation}
g_{1}\left( S_{0},S_{0}\right) +g_{2}\left( S_{0},S_{0}\right) =0.
\label{disclination_order}
\end{equation}
Now we consider the particular models.
\subsection{Full free energy expansion}
A widely used formula for $\sigma $ is Landau-de Gennes': 
\begin{equation}
\sigma _{LG}= a\rmTr\bfQ^{2}-b\rmTr
\bfQ^{3}+c\left[\rmTr\bfQ^{2}\right]^{2} ,  \label{free_energy}
\end{equation}
where $a$ is assumed to depend linearly on the temperature, 
whereas positive constants $b$, $c$ 
are considered temperature independent. 
For this free energy,
the \emph{uniaxial nematic} state is stable when $b^{2}>24ca$ with
the degree of orientational ordering 
\begin{equation}
S_{b}=\frac{b}{8c}\left( 1+\sqrt{1-\frac{64ca}{3b^{2}}}\right) .
\label{bulk_order}
\end{equation}
Taking into account the parametrization (\ref{eq_parametrization}) the suitably 
rescaled potential (\ref{free_energy}) may therefore be rewritten as
\begin{eqnarray}
\nonumber
\sigma_{LG} = k \left\{ S^{2}\left( 
\frac{1}{2}S^{2}-\frac{2}{3}S\left( S_{b}+S_{u}\right) +
S_{b}S_{u}\right) \right. \\
+ \left. \alpha^{2}\left[3S_uS_b+6(S_u+S_b)S+3S^{2}+\frac{9}{2}\alpha^{2} \right]
\right\} ,
\label{LG_free_energy}
\end{eqnarray}
so that the turning points for the uniaxial nematic phase 
with $\alpha=0$ occur
at $S=S_u, S_b$. 

The functions $g_{1},g_{2}$ then read 
\begin{eqnarray}
\nonumber
\lambda^{2} g_{1} &=&S_{u}S_{b}S+\left( S_{u}+S_{b}\right) 
\left\{ 3\alpha^{2}-S^{2}\right\} 
+S\left\{ 3\alpha ^{2}+S^{2}\right\} , \\
\nonumber
\lambda^{2}g_{2} &=&3S_{u}S_{b}\alpha 
+6\left( S_{u}+S_{b}\right) S\alpha +3\alpha
\left\{ 3\alpha ^{2}+S^{2}\right\} ,  
\end{eqnarray}
where $\lambda^{2}= \frac{9}{4}\kappa/k$ is the characteristic length.

The equation (\ref{disclination_order}) provides us with the values of the
order parameter and biaxiality on the disclination line:
\begin{equation}
\label{order_LG}
S_{0}=\alpha _{0}=-\frac{1}{2}S_{u} \:.
\end{equation}
\subsection{Biscari and Virga approach}
In order to obtain analytical expressions for the order parameter profile,
Biscari and Virga 
\cite{biscari.p:1997.a} 
used a quadratic approximation to
the full free energy $\sigma $, expressed in terms of $S$ and $\alpha$: 
\begin{equation}
\sigma _{BV}=k\left\{ \eta \left( S-S_{b}\right) ^{2}+\alpha ^{2}\right\} 
\label{BV_free_energy}
\end{equation}
with the following expressions for the functions $g_{1},g_{2}$: 
\[
g_{1}=\frac{\eta }{\lambda ^{2}}\left(S-S_b\right) \:,
\qquad
g_{2}=\frac{1}{\lambda ^{2}}\alpha \:. 
\]
The order parameter at the center of the disclination line then reads 
\begin{equation}
\label{order_BV}
S_{0}=\frac{\eta }{1+\eta }S_{b} \:.
\end{equation}
\subsection{Mottram and Hogan approach}
Another model, 
considered by Mottram and Hogan 
\cite{mottram.nj:1997.a} 
uses a quartic potential in $S$ 
but retains a quadratic potential in $\alpha $ 
\begin{equation}
\sigma_{MH} = k \left\{ \eta S^{2}\left( 
\frac{1}{2}S^{2}-\frac{2}{3}S\left( S_{b}+S_{u}\right) +
S_{b}S_{u}\right) +\alpha^{2}  \right\}
\label{MH_free_energy}
\end{equation}
This potential provides the following functions $g_{1},g_{2}$:
\[
g_{1}=\frac{\eta }{\lambda^{2}} S\left( S-S_{u}\right) 
\left(S-S_{b}\right) \:,
\qquad
g_{2}=\frac{1}{\lambda ^{2}}\alpha \:, 
\]
and the following value of the order parameter on the disclination line: 
\begin{equation}
\label{order_MH}
S_{0}=\frac{1}{2}\left\{ S_{u}+S_{b}\pm \sqrt{\left( S_{u}-S_{b}\right)
^{2}-4\eta ^{-1} }\right\} 
\end{equation}
\subsection{Maier-Saupe approach}
Another mean-field approach to the description of the disclination line was
considered by Sigillo et al
\cite{sigillo.i:1998.a}.
They applied Maier-Saupe theory and obtained
expressions for the functions $g_{1},g_{2}$.
The Maier-Saupe theory has an intermolecular potential strength $U$ as
the only adjustable parameter, determining the correct value of the bulk order
parameter $S_{b}\left( U\right) $. The value of the order parameter 
$S_{0}$ on the disclination line is then fixed and determined by the value of 
$S_{b}$.
\section{Molecular model and simulation methods}
\label{sec:model}
The molecules in this study are modelled as
hard ellipsoids of revolution of elongation $e=A/B=15$,
where $A$ is the length of the major axis 
and $B$ the length of the two equal minor axes.
The phase diagram and properties of this family of models
are well studied
\cite{frenkel.d:1984.a,frenkel.d:1985.a,1993.g,1993.j,1992.a}.
Units of length are chosen such that
$AB^2=1$,
making the molecular volume equal to that of a sphere of unit diameter.
It is useful to express the density as a fraction of the close-packed
density for perfectly aligned hard ellipsoids,
assuming an affinely-transformed face-centred cubic or hexagonal close
packed lattice;
in reduced units $\varrho_{\rm cp}^*=\varrho_{\rm cp}AB^2 = \sqrt{2}$.
Henceforth the asterisks denoting reduced quantities will be omitted.

The molecules were confined within a cylinder of radius $R$,
and height $H$, with periodic boundary conditions applied
in the $z$ direction (the symmetry axis).
The cylindrical confining walls
are defined by the condition that they
cannot be penetrated by the \emph{centres} of the ellipsoidal molecules.
Packing considerations generate \emph{homeotropic} ordering at the surface,
i.e.\ the molecules prefer to orient normal to the interface,
without the need to apply an explicit ordering field.
The properties of such surfaces in planar geometry
were investigated previously
\cite{1999.b}.
For these particles,
the isotropic-nematic phase transition occurs at
quite low density,
$\varrho/\varrho_{\rm cp}\approx0.2$.
Temperature is not a significant thermodynamic quantity in this model;
we set $k_{\rm B}T=1$ throughout.

Monte Carlo simulations were carried out for systems 
of the following cylinder radii:
$R/A = 2.08, 2.67, 4.00, 5.33$.
The corresponding numbers of ellipsoids were
$N=3500, 6000, 13000, 22000$.
The height $H/A$ was in the range 2.6--2.8 in all cases.
All the simulations were conducted at the
same state point used in the earlier study
\cite{1999.b},
corresponding to a bulk pressure
$P=2.0$ in the above reduced units;
this is significantly higher than the 
isotropic-nematic coexistence pressure $P\approx1.49$
\cite{1996.f}.
No external fields were applied,
and conventional Monte Carlo moves were employed,
with translational and rotational displacements chosen to give
a reasonable acceptance rate.

The simulation results were analyzed to give
a density profile,
averaged over cylindrical shells of width $0.125B$--$0.25B$,
and an order tensor profile obtained by averaging 
\[
Q_{\alpha\beta}(\rho_i) =  \left\langle \frac{1}{n_i} \sum_{j=1}^{n_i}
\left\{\frac{3}{2}u_{j\alpha} u_{j\beta} 
- \frac{1}{2}\delta_{\alpha\beta} \right\} \right\rangle
\qquad \alpha,\beta=\rho,\theta,z
\]
where there are $n_i$ molecules present in bin $i$;
$\delta_{\alpha\beta}$ is the Kronecker delta.
The axis system is resolved as before into 
radial ($\rho$), 
tangential ($\theta$), and 
axial ($z$) components,
for the purposes of accumulating these functions.
Therefore, the tensor components $Q_{\alpha\beta}$
have been averaged over global rotations of the system
(positions and orientations together)
about the symmetry axis,
as well as global translations in the $z$ direction.
Diagonalizing this tensor, for each bin,
gives three eigenvalues and three corresponding eigenvectors.
This procedure allows us to test for 
(a) the planar radial structure,
for which the eigenvectors lie along the
$\rho,\theta,z$ coordinate axes, independent of $\rho$,
and the eigenvalues are
$Q_{\rho\rho}(\rho)$, $Q_{\theta\theta}(\rho)$, $Q_{zz}(\rho)$);
and 
(b) the escaped radial structure,
for which one of the eigenvectors
points along the $\theta$ direction for all values of $\rho$,
and the other two lie in the $\rho,z$ plane,
changing their orientation as $\rho$ varies.

The above tensor is not sensitive to any breaking of
axial symmetry: instead, it gives a cylindrical average.
To give a full representation of the positional variation
of the orientational order,
we must retain the full spatial dependence of $\bfQ$.
We find it convenient to calculate
\begin{equation}
\label{eqn:Stensor}
S_{\alpha\beta}(\bfr) = \frac{1}{n_{\calR}} \sum_{j\in\calR} 
u_{j\alpha} u_{j\beta}
\qquad \alpha,\beta=x,y,z
\end{equation}
where the sum is conducted over 
the $n_{\calR}$ molecules found to be in a neighbourhood $\calR$
of the chosen point $\bfr$.
The eigenvectors of this $\bfS$ tensor are the same as those
of the corresponding $\bfQ$;
the eigenvalues are linearly related
and are non-negative,
so the tensor may be visually represented as a spheroid,
whose principal axis directions and lengths
are given by the eigenvectors and corresponding eigenvalues
\cite{1989.a}.
Then, a well-ordered uniaxial region appears as an elongated,
prolate, shape; the uniaxial core of a disclination defect of strength
$+1$ would appear as an oblate spheroid with its symmetry axis along
the axis of the cylinder; disordered regions would correspond to a
spherical shape, and so on.
\section{Simulation results}
\label{sec:results}
\begin{figure}
\includegraphics[width=8cm]{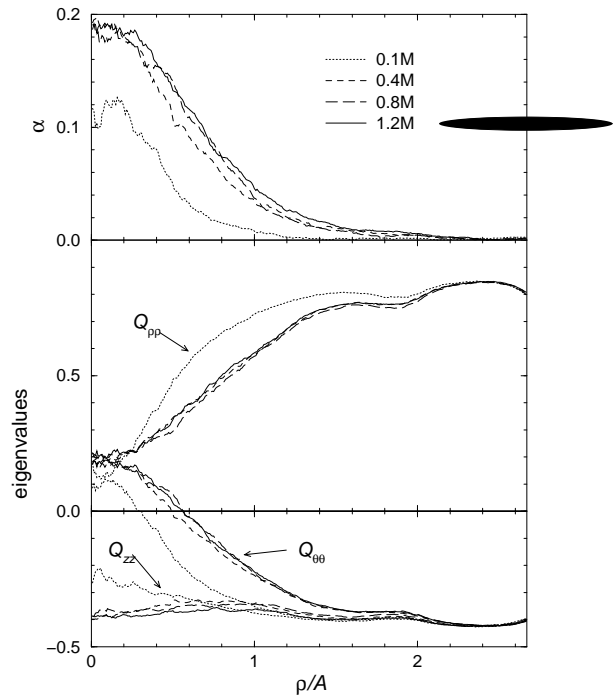}
\caption[Fig 1]{\label{fig:1} 
Time evolution of system started from perfect planar radial configuration, 
in cylinder of radius $R/A=2.67$.
We show order tensor eigenvalue profiles averaged over
$10^5$ sweeps, taken at the indicated times
(units of $\times10^6$ sweeps).
The top eigenvalue is $Q_{\rho\rho}$,
the middle one $Q_{\theta\theta}$
and the bottom one $Q_{zz}$ in each case.
In the upper panel we plot the biaxiality parameter
$\alpha = \frac{1}{3}(Q_{\theta\theta}-Q_{zz})$.
Note that the biaxial core develops quite rapidly,
and that equilibration is essentially complete
between 0.4 and 0.8 $\times10^6$ sweeps.
The final equilibrium configuration is identical with that obtained
from a disordered starting configuration.
The shaded ellipse indicates the outermost, homeotropically
anchored, layer at the cylinder wall.
}
\end{figure}

Most results were obtained from
starting configurations in which the ellipsoids were
positioned randomly, avoiding overlaps, but 
perfectly aligned in the planar radial configuration.
Typical equilibration times for these systems were around
$7\times10^5$ MC sweeps 
(one sweep is one attempted move per particle),
and this was monitored through the orientation profiles defined above.
The development of a biaxial core in the system with $R/A = 2.67$
is shown in Fig.~\ref{fig:1};
the other systems evolved in a similar way.
Following equilibration, production runs of approximately
$5\times10^5$ MC sweeps were undertaken.

\begin{figure}
\includegraphics[width=8cm]{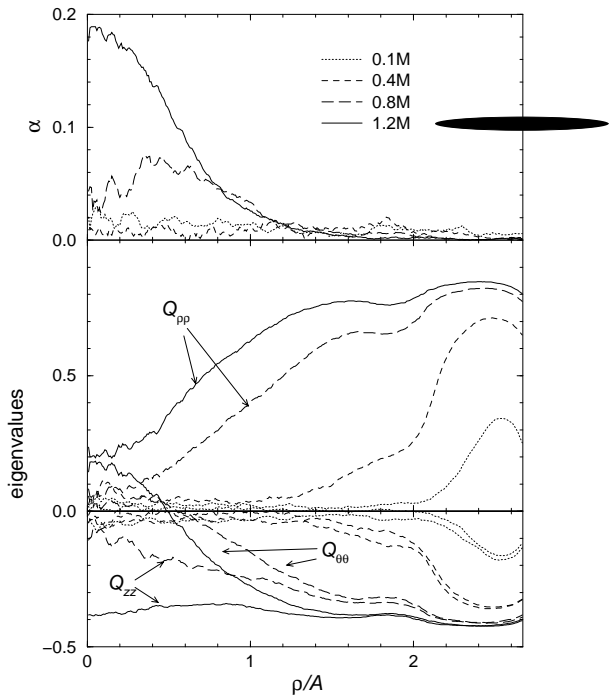}
\caption[Fig 2]{\label{fig:2} 
Time evolution of system started from orientationally disordered configuration, 
in cylinder of radius $R/A=2.67$.
Notation as for Fig.~\protect\ref{fig:1}.
Note that the ordering remains uniaxial at the boundary,
and propagates in towards the centre;
as the ordering reaches the centre,
the biaxial disclination core develops.
The final equilibrium configuration is identical with that obtained
from a planar radial starting configuration.
}
\end{figure}

Checks for convergence were carried out,
starting from configurations in which molecular orientations
as well as positions were disordered.
The time evolution of the $R/A = 2.67$ system 
is shown in Fig.~\ref{fig:2}.
Note that the ordering remains uniaxial at the boundary,
and propagates in towards the centre;
as the ordering reaches the centre,
the biaxial disclination core develops.
An equilibration period of
$10^6$ sweeps was needed to reach the equilibrium planar radial structure,
and a subsequent production run of $2\times10^5$ sweeps yielded identical
results to those obtained from the planar radial starting point.
For the larger cylinder radii, similar behaviour was observed,
but the timescale for propagation of the order from the boundary
to the centre was correspondingly longer.
In each case,
a run was conducted
which was of sufficient length
to confirm that the correct structure was becoming established.

In no case was the escaped radial structure formed during these runs.
We have carried out some preliminary tests,
for the pores of smaller radius, $R/A = 2.08, 2.67$,
in which the escaped radial
structure was stabilized by the application of a uniform orienting field
favouring alignment along the cylinder axis.
Following removal of this field, the planar structure was
seen to be recovered
on a simulation timescale of $5\times10^5$ sweeps.
Thus, the planar structure is thermodynamically stable for these cases,
and it appears to be 
at least metastable on the timescales of our simulations
for the larger pores.

\begin{figure}
\includegraphics[width=8cm]{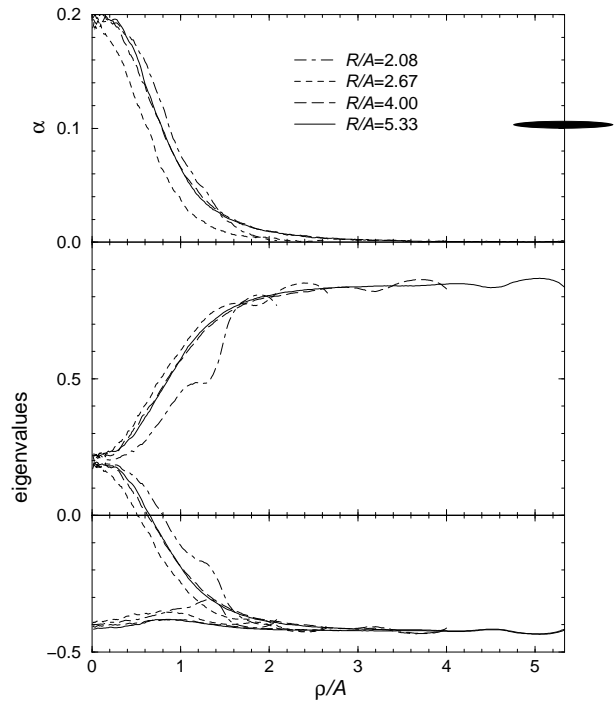}
\caption[Fig 3]{\label{fig:3} 
Equilibrium order tensor eigenvalue profiles
for cylinders of indicated radii. 
For the smallest two cylinders, the walls clearly have an effect
on the defect core structure;
the results for the two largest cylinders are essentially 
indistinguishable.
Notation as for Fig.~\protect\ref{fig:1}.
}
\end{figure}

In Fig.~\ref{fig:3}
the order parameter profiles are presented
for all the different cylinder sizes.
The biaxiality profile
$\alpha(\rho) = \frac{1}{3}(Q_{\theta\theta}-Q_{zz})$
indicates the extent of the core region.
For the smallest two cylinders,
it is clear that the walls act to deform the disclination core.
For $R/A=2.08$, the core region is enlarged,
while for $R/A=2.67$ it is compressed.
The results for the two largest cylinders are almost 
indistinguishable.

\begin{figure}
\includegraphics[width=8cm]{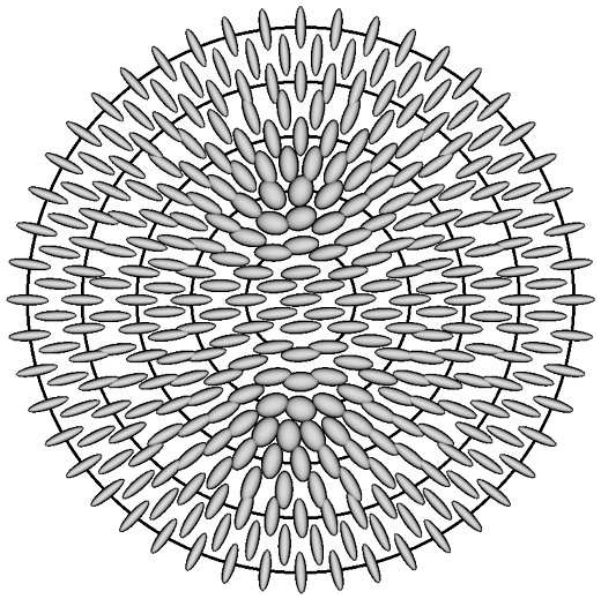}
\caption[Fig 4]{\label{fig:4} 
Distribution of order tensor in the plane perpendicular to
the cylinder axis, in the core region, for the case $R/A = 4.00$.  
Concentric circles at half-integer intervals of radius $\rho/A$
are plotted as a guide.  The spheroid sizes are not physically significant.
The figure is oriented so that the $+1/2$ defects lie above and below
the cylinder axis.
}
\end{figure}

In Fig.~\ref{fig:4}
we show the order tensor variation in the $xy$ plane
(no variation with $z$-coordinate was observed)
for the core region in the $R/A = 4.00$ case.
The $\bfS$-tensor spheroid of eqn~(\ref{eqn:Stensor})
is plotted at a number of points
$(x_s,y_s)$
which lie on circles centred on the cylinder axis, separated by $0.25A$,
with the neighborhood $\calR$ of each point defined to include all molecules
whose centres $(x,y,z)$ satisfy $\sqrt{(x-x_s)^2+(y-y_s)^2}<0.25A$.
The tensors are averaged over a run of $10^5$ sweeps.
The figure indicates that the core is actually best described
as a pair of disclinations each of strength $+1/2$, 
symmetrically arranged at $\rho/A\approx$1--1.25
from the cylinder axis.
Very similar results are seen
for all the three largest pores, $R/A = 2.67, 4.00, 5.33$,
while for the smallest pore,
$R/A=2.08$, the defects are slightly further out, 
at $\rho/A\approx 1.5$.
There is a small region of almost unperturbed
uniaxial nematic liquid crystal
around $\rho\approx0$,
with the director perpendicular to the cylinder axis.
We note that similar defect pairs are seen in the
two-dimensional simulations reported in Ref.~\cite{Dzubiella:1999.a}.

The order tensor profiles reported in Fig.~\ref{fig:3}
are properly regarded
as axial averages of structures which are not themselves
axially symmetric on the timescales of the simulation.
This is why the lowest eigenvalue in the profiles
of Fig.~\ref{fig:3} adopts similar values inside and outside the core:
it corresponds to the eigenvector along the cylinder axis,
and is unaffected by the axial averaging.
We discuss in section \ref{sec:conclusions}
the validity of carrying out such an axial average,
simply noting here that it is the most straightforward way of
comparing our simulation results with theories
which assume cylindrical symmetry.

\begin{figure}
\includegraphics[width=8cm]{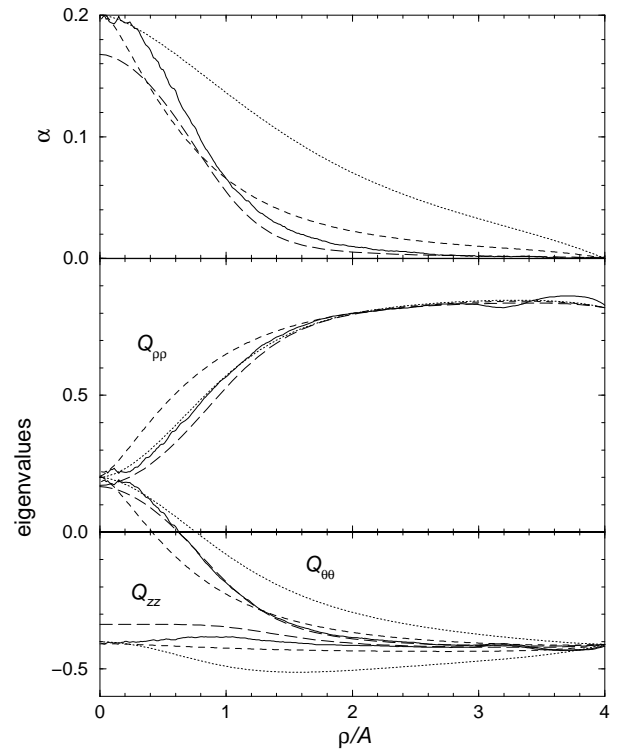}
\caption[Fig 5]{\label{fig:5} 
Fits of simulation results (solid lines) to theoretical predictions
discussed in the text. 
The best fit in each case was obtained for the following parameters:
Mottram, Hogan \protect\cite{mottram.nj:1997.a} (dotted lines):
 eqn~(\ref{MH_free_energy}), with
 $\eta /\lambda ^{2}=1.58$, $\lambda^{-2}=1$, 
 $R=6.5$, $S_{b}=0.88$ $S_{s}=0.82$, $S_{u}=0.01.$
Biscari, Virga \protect\cite{biscari.p:1997.a} (dashed lines):
 eqn~(\ref{BV_free_energy}), with
 $\eta /\lambda ^{2}=1$, $\lambda ^{-2}=3.4$,
 $R=6$, $S_{b}=0.88$, $S_{s}=0.82$.
Sigillo, Greco, Marrucci \protect\cite{sigillo.i:1998.a} (long dashed lines):
  $U=13$, $S_{s}=0.82$, $R=7$.
Notation as for Fig.~\protect\ref{fig:1}.
}
\end{figure}

\begin{figure}
\includegraphics[width=8cm]{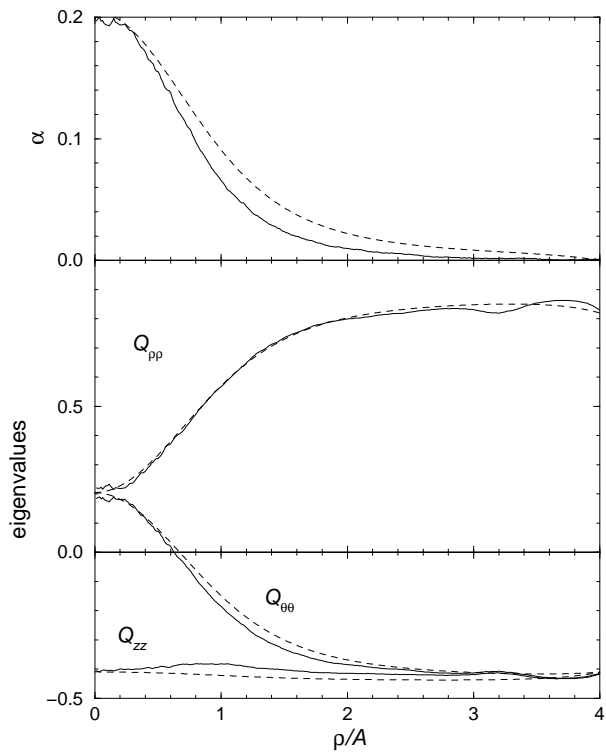}
\caption[Fig 6]{\label{fig:6} 
Fit of simulation results (solid lines) to the Landau-de Gennes
free energy expansion (\ref{LG_free_energy})
using the approach of 
Schopohl and Sluckin \protect\cite{schopohl.n:1987.a,schopohl.n:1988.a}
(dashed lines). 
The best fit in each case was obtained for
 $\lambda ^{-2}=1.42$,
 $R=6$, $S_{b}=0.88$, $S_{s}=0.82$, $S_{u}=-0.4$.
Notation as for Fig.~\protect\ref{fig:1}.
}
\end{figure}

We have fitted the results for $R/A=4.00$
using the phenomenological theories of section~\ref{sec:theory}.
In each case,
the boundary-value problem of
eqns~(\ref{general_equations}), (\ref{boundary_0}), (\ref{boundary_R}) ,
was solved using the \emph{relaxation method} 
\cite{press.wh:1992.a}. 
In Figs~\ref{fig:5}, \ref{fig:6}, 
the fitting curves and the original simulation data are shown. 
Note, that to fit the simulation data we first determind the 
value of the parameter
$\eta$ or $S_u$ using 
equations (\ref{order_LG}), (\ref{order_BV}), (\ref{order_MH}).
Then we performed suitable rescaling, changing the factor $\lambda$.

\section{Discussion}
\label{sec:conclusions}

One can see in Figs~\ref{fig:5}, \ref{fig:6}, 
that the slope of the fitting curves in general reflects the
structure of the core: 
the center of the core is strongly biaxial, extending over
a few units of length, 
as shown by the splitting of the eigenvalues 
$Q_{\theta\theta}$, $Q_{zz}$,
and hence the non-zero biaxiality parameter $\alpha(\rho)$.

At the same time, the difference between the descriptions is also evident.
Biscari and Virga's theory gives an incorrect shape of 
$S\left(\rho\right)$ 
for small values of $S$ (and hence, small values of $\rho$ here). 
This is fairly predictable,
since a quadratic expansion of the free energy density 
$F\sim \eta \left(S-S_{b}\right) ^{2} $ 
was used, which is valid only for small deviations of
the order parameter from the \emph{bulk} value $S_{b} $.

The more sophisticated form of the free energy 
used by Mottram and Hogan (up to fourth order in $S$) 
corrects the slope of the curve for the order
parameter $S=Q_{\rho \rho } $. However, the \emph{biaxial} part, 
$Q_{\theta\theta }-Q_{zz} $, 
still has only qualitative agreement with the simulation
results, which is probably because of the
\emph{quadratic} approximation to the
biaxial part of the free energy.
This is particularly apparent in the $\alpha(\rho)$ profile,
in which the biaxiality extends far beyond the core radius
which would be deduced from $Q_{\rho\rho}$.

In spite of having only one adjustable parameter, 
the Maier-Saupe theory of Sigillo et al.\
gives a very
realistic description of the disclination core structure. However, it
predicts slightly lower core biaxiality
than obtained in the simulations.

The most accurate fitting we obtained used the full expansion of the free
energy in powers of the order parameter tensor (\ref{free_energy}),
as shown in Fig.~\ref{fig:6}.
This form, originally used by Schopohl and Sluckin,
manages to reproduce the overall extent of the biaxial region quite well,
while fitting the limiting behaviour of both $\alpha(\rho)$ and $S(\rho)$
as $\rho\rightarrow0$,
and the magnitude of $S$ in the bulk.
Therefore, we can conclude, that all the terms in the free energy expansion
(or, at least up to the fourth order in the order parameter tensor) should
be taken into account if one want to make the correct \emph{quantitative}
description of the disclination core structure.

Fig~\ref{fig:4} shows that the core structure seen on the simulation
timescale is not, in fact, cylindrically symmetrical,
and this raises questions regarding the validity of averaging the
order tensor over rotations about the cylinder axis.
Such averaging may arise in a natural way.
We see a small amount of rotation of the defect pair around the axis,
of the order of 10--15 degrees,
as the core structure evolves in time,
during the $5\times10^5$-sweep simulation runs conducted here.
If we roughly equate this to a real-time period of the order of nanoseconds,
it seems possible that significant rotation will occur on the 
experimental timescale.
The situation is further complicated when one considers correlations along
the $z$-direction; the periodic box employed here is quite small,
but use of a much longer cylindrical pore might reveal some 
twisting of the separate $+1/2$ defect lines about the cylinder axis.
Both effects would result in the kind of cylindrical averaging 
which we have carried out,
but we can currently only speculate on this.

A second question concerns the
distance between the $+1/2$ defect lines.
We do not observe separation of the defect pair, 
either as the simulations proceed in time,
or as we study larger pores.
Nonetheless, it is quite possible that the pore walls are
confining the defect pair to the vicinity of the axis,
and that they might separate if a much larger system were
simulated.
We note that, if they were to completely separate and approach the
pore walls,
the result would be the planar polar structure discussed elsewhere
\cite{crawford.gp:1991.a,crawford.gp:1992.a}.
It would be interesting to calculate free energies
as a function of the defect separation to investigate this.

Finally,
we may expect the escaped radial structure to become increasingly favoured
as the pore radius increases;
it is clearly not the most stable structure for the smallest
pores studied here.
Once more, free energy calculations are needed to make a proper comparison.
Further work on these aspects is in progress.
\section*{Acknowledgements}
The advice of Nigel Mottram
is gratefully acknowledged,
and the referee's comments on the first version of this paper
were very helpful.
This research was supported by EPSRC,
and by the University of Bristol Department of Physics. 
DA also acknowledges 
support through the grants PSU082002 of the 
International Soros Science Education Program and 
ORS/99007015 of the Overseas Research Students Award.
The authors of Ref.~\cite{Dzubiella:1999.a} 
are thanked for providing a preprint.
\bibliography{journals,main,mike,extra} 
\end{document}